# Journal Name

## ARTICLE

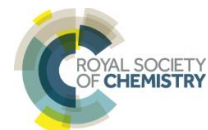

# Nanofabrication inside Microfluidic Chips for Biomimetic Wet-Spinning of Fibres

Jonas Lölsberg[a,b,+], John Linkhorst[b,+], Arne Cinar[b], Alexander Jans[a], Alexander J. C. Kuehne[a] and Matthias Wessling[a,b,*]

Microfluidics is an established multidisciplinary research domain with widespread applications in the fields of medicine, biotechnology and engineering. Conventional production methods of microfluidic chips have been limited to planar structures, preventing the exploitation of truly three-dimensional architectures for applications such as multi-phase droplet preparation or wet-phase fibre spinning. Here the challenge of nanofabrication inside a microfluidic chip is tackled for the showcase of a spider-inspired spinneret. Multiphoton lithography, an additive manufacturing method, was used to produce free-form microfluidic masters, subsequently replicated by soft lithography. Into the resulting microfluidic device, a three-dimensional spider-inspired spinneret was directly fabricated in-chip via multiphoton lithography. Applying this unprecedented fabrication strategy, the to date smallest spinneret nozzle is produced. This spinneret resides tightly sealed, connecting it to the macroscopic world. Its functionality is demonstrated by wet-spinning of single-digit micron fibres through a polyacrylonitrile coagulation process induced by a water sheath layer. The methodology developed here demonstrates fabrication strategies to interface complex architectures into classical microfluidic platforms. Using multiphoton lithography for in-chip fabrication adopts a high spatial resolution technology for improving geometry and thus flow control inside microfluidic chips. The showcased fabrication methodology is generic and will be applicable to multiple challenges in fluid control and beyond.

## Introduction

Biomimicry in textiles has a long history of providing man with ever-growing knowledge about the production of artificial fibres.[1] In particular, spider silk fibres attract attention with their superior strength to weight ratio.[2-5] Dozens of potential applications are proposed for spider silk fibres in the domain of regenerative medicine, for example as surgical sutures or as artificial skin scaffolds.[6] However, the bulk production of artificial spider silk fibres with native dimensions of smaller than 5 μm diameter remains a bottleneck.[7-13]

Several promising microfluidic approaches were used in the past to generate functional fibres of small dimensions. For example, coaxial liquid flow (flow-focusing) devices were applied to produce anisotropic hollow fibres by *in situ* photo-polymerization or soft hydrogel fibres for the use as degradable scaffolds. Flow-focusing microfluidic devices were utilized for the production of cell adhesive porous/non-porous Janus fibres using photo-curable polyurethane, for phase separation spinning of biodegradable tissue, for wet-phase spinning of silk fibres and for self-assembling recombinant spider silk fibres.[14-21] These microfluidic wet-spinning processes yield small fibres within the double-digit micron scale as measured in a wet state.

Today's microfluidic fibre syntheses with controlled shape and surface topology are hampered by the planar environment of current microfluidic setups produced by soft-lithography. The lateral dimensions of these microfluidic channels are often limited to a range of hundreds rather than tens of micrometres, and it is through the use of special techniques such as casting-shrinkage methods to decrease the dimensions of microfluidic channels to the double and single digit micron scale.[22]

A potential approach to fabricate such small fibre dimensions may be the integration of a miniaturized spinneret into a microfluidic chip. A potential technology for embedding micro- and nanoscale systems into microfluidic devices is multiphoton lithography, a direct additive manufacturing technique to produce three-dimensional polymeric structures, such as tiny flow channels or periodic grating structures on the order of the wavelength of visible light.[23-27] Multiphoton lithography is used for in-chip fabrication of embedded patterns, filters, valves, mixers, and sensors.[28-41] However, interfacing fine fluidic structures with macroscopic flow

[a.] DWI - Leibniz Institute for Interactive Materials, Forckenbeckstr. 50, 52074 Aachen, Germany

[b.] RWTH Aachen University, AVT - Chemical Process Engineering, Forckenbeckstr. 51, 52074 Aachen, Germany

[+] These authors contributed equally to this work.

[*] Phone: +49-241-80 95470, Manuscripts.cvt@avt.rwth-aachen.de

[†] Electronic supplementary information (ESI) available: In-chip laser lithography fabrication of the embedded spinneret in Vid. S1; Sealing the embedded spinneret in Vid. S2; In-chip fibre synthesis in Vid. S3; Stereolithography file for the microfluidic master *master_V5.stl*; Stereolithography file of the embedded spinneret *nozzle_V8.stl*; See DOI: 10.1039/x0xx00000x

channels with inlet- and outlet connections still remains a challenge as to our knowledge pressure resistant and liquid tight connections are not reported.

We therefore hypothesize that microfluidic devices in conjunction with in-chip fabrication methods could be a new road to miniaturize artificial spider spinneret nozzles generating fibres with native dimensions in a controlled environment. To support the hypothesis, we use multiphoton lithography as a tool to embed a significantly smaller nozzle by in-chip laser lithography into a larger flow channel fabricated by dip-in laser lithography. This unprecedented fabrication process integrates dip-in laser lithography, soft lithography, and in-chip laser lithography to fabricate an integrated micro-spinneret nozzle within a microfluidic device, which would have been unachievable using contemporary fabrication techniques. This device is utilized to produce thin fibres in a wet-spinning process where a polymer solution made of polyacrylonitrile (PAN) is coagulated and solidified into a solid microfibre with single digit micrometre dimensions.

## Results and discussion

To integrate the micro-spinneret into the geometry of a microfluidic channel, we design a novel production pathway, which is summarized by the following three steps: The fabrication of free-form microfluidic masters using dip-in laser lithography (Fig. 1a), soft-lithography to mould a channel geometry from the positive master relief into poly(dimethylsiloxane) (PDMS) (Fig. 1b) and finally in-chip laser lithography for embedding the spinneret inside the microfluidic channel (Fig. 1c).

### Fabrication of free-form microfluidic devices

The output of the dip-in laser lithography and soft-lithography methods is a free-form fabricated microfluidic PDMS chip (Fig. 2b) bond to a transparent glass substrate. The microfluidic chip consists of two inlet channels and one outlet channel. The openings of the inlet and outlet channels are structured with a zigzag pattern, a so-called labyrinth seal, to facilitate fluid-tight and pressure-resistant interfacing between the PDMS channel and the embedded spinneret.[42] There are two additional channels perpendicular to the outlet channel, designed to be filled with sealant, improving adhesion and structural integrity. The five channels merge at a junction, the nozzle holder. The channel sidewalls of the junction have an angle of 30° widening towards the plasma-bond transparent glass slide, which is an essential pre-requisite to prevent shadowing effects when using in-chip laser lithography.[43] Shadowing effects can arise by absorption, reflection and diffraction of the laser beam due to refractive index differences between the PDMS channels, the polymerized structures and the unpolymerized photoresist. Such effects distort and deflect the focal point of the laser, leading to insufficient power to initiate the two-photon polymerization reaction. Preventing shadowing is therefore the key to tight connections between the microfluidic chip and the embedded spinneret.

### In-chip fluid dynamics simulations

Prior to the in-chip fabrication process of the spinneret, we model the fluid dynamic behaviour (Fig. 3) and we take inspiration from the spinning process of spider silk. During the silk spinning process, the spider produces fibres with an outlet velocity on the order of 1 cm s$^{-1}$ at the nozzle tip of its spin gland.[2, 3, 44] To generate a homogenous flow profile with equal flow velocity of bore and sheath fluid, we set the bore volume flow rate to 4.1 µL s$^{-1}$ and the sheath flow rate to 150.7 µL s$^{-1}$ according to the ratio of lumen cross-section and sheath flow channel cross section. From the inlet channels of the sheath fluid, we observe a decelerating flow due to the increased channel diameter when the channel opens into the nozzle holder that contains the spinneret nozzle (Fig. 3a). The bore fluid supplied by the other inlet channel accelerates towards the end of the spinneret due to the tapered nozzle (Fig. 3b). This flow configuration induces shear forces on the bore fluid, which further increase toward the nozzle tip. The bore fluid diameter is further reduced when leaving the nozzle as it is focused by the surrounding sheath flow. The modelled coaxial flow distribution inside the spinneret is presented at five discrete positions along the flow path (Fig. 3c). The in-plane velocity distribution develops towards complete homogeneity from the nozzle holder inlet to the nozzle opening. The final optimized spinneret geometry (Fig. 3d) is then used for the in-chip lithography fabrication step.

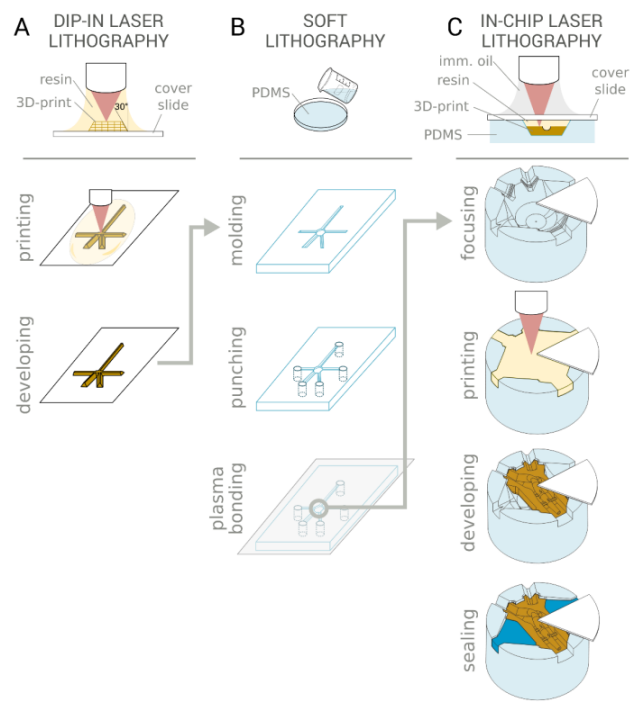

**Fig. 1**    Fabrication process for integrating a spinneret into microfluidic channels. (a) Maskless dip-in laser lithography used to print free-form positive moulds with an internal rectangular scaffold for structural integrity and an external tilted wall. (b) Mould replication by soft lithography using PDMS. (c) In-chip laser lithography to fabricate an embedded spinneret into the nozzle holder of the microfluidic PDMS chip.

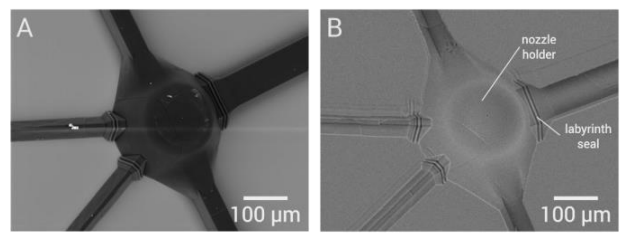

**Fig. 2**    SEM images of (a) the free-form positive mould printed by dip-in laser lithography and (b) the PDMS mould replicated by soft lithography. In total, five tilted channels merge into a junction, the nozzle holder, where the spinneret will be embedded.

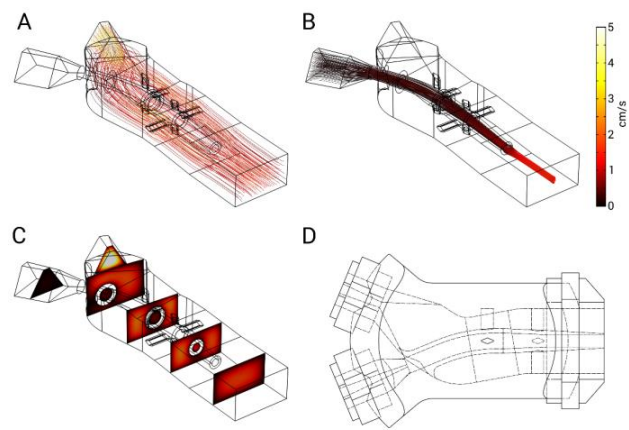

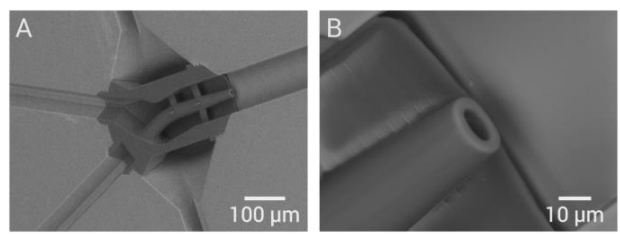

**Fig. 3** Simulation of the flow field inside the embedded spinneret visualized by (a) streamlines of the outer surrounding sheath flow regime and (b) streamlines of the inner bore flow regime. (c) The lateral distribution of the sheath flow is shown at five discrete positions along the flow path. The streamlines and planes are colour-coded by velocity magnitude. CAD drawing of the optimized geometry from the (d) top view.

**Fig. 4** SEM images of (a) the embedded spinneret interlocked into a microfluidic channel and (b) showing a close-up of the 12 μm nozzle tip. A video showing the printing process can be found in the ESI Vid.S1†.

**In-chip laser lithography fabrication of the embedded spinneret**

Printing the optimized nozzle into the nozzle holder is performed by filling it with negative tone photoresist and subsequent initiation of the printing process. The laser light exposure induces a two-photon polymerization and crosslinking reaction, curing and hardening the resist material in the desired three-dimensional geometry (see ESI Vid. S1†). After completion, the inlet and outlet channels are used to flow in a solvent for removal of unpolymerized photoresist, revealing the delicate spinneret geometry with a nozzle diameter of 12 μm (Fig. 4). The resulting spinneret geometry is perfectly aligned with regards to the sidewalls of the junction and the surrounding microfluidic channel network. The inner spinneret nozzle is centred and mechanically stabilized using fins reaching towards the sheath flow channel sidewalls. The fins are designed to avoid any influence of the flow profile.

Additional sealing with silane-based glue from the perpendicular side channels encloses the printed spinneret geometry and creeps into small voids to block all potential leakages. After curing the glue (see ESI Vid. S2†), a liquid-tight operation up to an inlet pressure of $2\times10^5$ Pa is possible. The surface quality of the embedded spinneret and the surrounding PDMS chip depends on the applied laser lithography printing parameters, including the slicing and hatching distances as well as the laser power and speed.[45, 46] At high laser power, we observe bubbles emerging at the PDMS and glass interfaces due to a laser light induced rapid heating and evaporation of the photo-acid. This problem is avoided by programming a spatial laser intensity gradient into the 3D model, reducing the power when the laser is in close proximity to the PDMS and glass interfaces.

**In-chip colour experiments**

The fluid dynamic behaviour of the embedded spinneret is first tested using coloured aqueous solutions as shown in the micrograph in Fig. 5a. The blue inner bore solution is surrounded by the yellow outer sheath solution. This simple testing method verifies the capability to focus the inner bore flow. The flow-focusing effect is proportional to the ratio of the applied sheath and bore pressures. When increasing the sheath pressure the diameter of the blue inner bore solution is reduced. This phenomenon is also observed when modelling the fluid dynamic behaviour of the spinneret. In addition, the test verifies leakage-free operation, proving there is intimate sealing of the microfluidic chip with the interfacing spinneret and the surrounding adhesive.

**In-chip fibre syntheses**

To highlight the applicability of our embedded spider inspired spinneret, we perform continuous flow-focused wet spinning of a single PAN fibre over an extended fabrication time of 15 h. During screening experiments, we adjusted the bore and sheath fluid compositions to find stable process conditions. The initial bore fluid contains 10 % (w/w) PAN dissolved in dimethyl sulfoxide (DMSO) resulting in a high bore pressure loss, which is reduced by decreasing the PAN concentration to 2 % (w/w). This adaptation allows continuous jetting of the bore fluid with an estimated speed of 5 cm s$^{-1}$ at $0.5\times10^5$ Pa inlet pressure and prevents fluid instabilities at the nozzle tip. To avoid immediate coagulation of the PAN bore fluid at the orifice of the spinneret, which would lead to nozzle fouling or blockage (see ESI Vid. S3†), we added 65 % (w/w) DMSO in deionized water as the sheath fluid delays the phase inversion significantly.[47] This adaption enables a more gradual PAN fibre coagulation in the microfluidic chip.[48] The flow-focusing effect is directly visible at the nozzle tip in the micrograph in Fig. 5b as the emerging PAN solution reduces its diameter from 12 µm at the nozzle orifice to 4 µm within the first 119 µm. As result of the stable flow-focusing effect, which is maintained by accurate pressure control, this wet-spinning process yields PAN fibres with diameters of 2.7 µm as shown in Fig. 6c and d. Such narrow diameter PAN fibres can usually only be obtained after stretching and elongation.[49] To our knowledge, micrometre-sized PAN fibres from solutions as low as 2 % (w/w) have never been directly obtained before, reflecting the high control and optimized spinning conditions inside the flow-focusing microfluidic device (see ESI Vid. S3†).[50] As the fibre is collected in the coagulation bath into which the outlet channel of the chip is immersed, we observe an unaligned fibre arrangement produced by flow deceleration (Fig. 6a). Other collection setups allow for the collection of the fibres in an aligned way (Fig. 6b).[13] We selected single fibres and applied the HDMS drying method to investigate the fibre morphology in a dry state by FESEM (Fig. 6c). The HMDS drying procedure has proven to be a fast and simple tool showing excellent results for drying fibres. The fibre topology appears to be corrugated as a result of the coagulation process, when the non-solvent diffuses from the fibre surface through the cross-section during solidification. We assume that this solvent/non-solvent exchange forms a rigid fibre skin before the centre of the fibre is solidified, yielding the corrugated surface topology.[51] This assumption is supported by the fact that the textures are aligned with the fibre direction, potentially resulting in a skin-core structure (Fig. 6d).[52] All examined fibres exhibit only small deviations in diameter further supporting excellent control and reproducibility of the PAN spinning conditions.

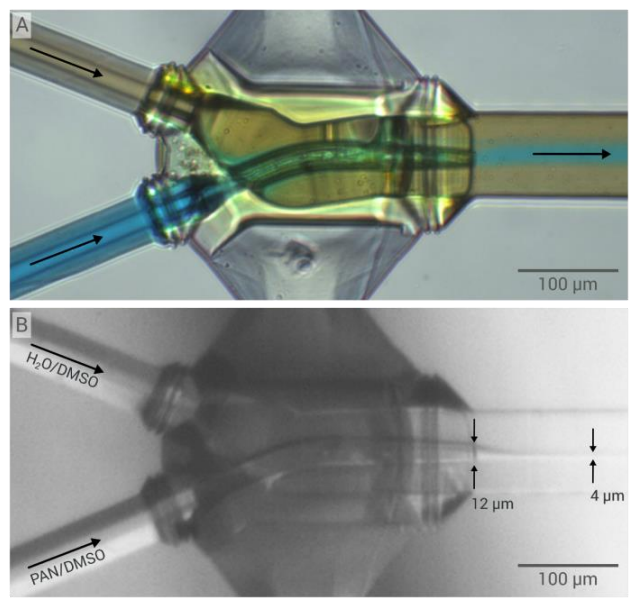

**Fig. 5** Optical micrographs of (a) the spinneret operated with an inner blue bore fluid and an outer surrounding yellow sheath fluid. The arrows represent the flow direction inside the microfluidic chip. (b) Fabrication of micrometre-sized PAN fibres in a wet-spinning process (image is a video still). The small arrows visualize the flow-focusing effect, which reduces the diameter of the emerging polymer solution coagulated in the water sheath layer. See the full video in the ESI Vid.S3†.

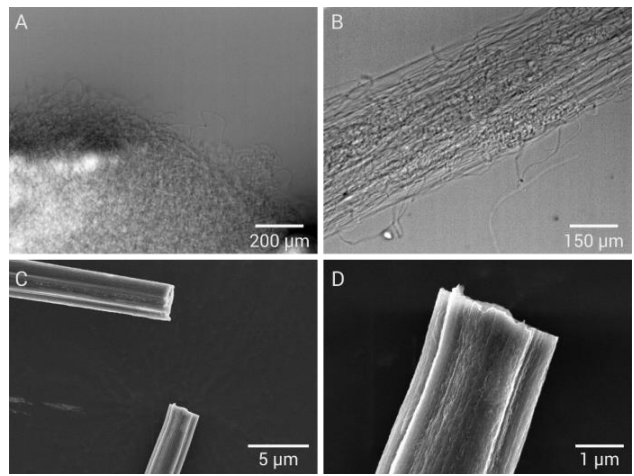

**Fig. 6** Optical micrographs of (a) the unaligned fibre arrangement and (b) aligned fibres after the wet-spinning process of PAN. FESEM images of (c) a spun fibre that has broken in half after solvent drying and (d) a higher magnification image showing the surface topology of an individual fibre with a diameter of 2.7 µm.

## Experimental

### Dip-in laser-lithography of free-form positive moulds

A schematic of the dip-in laser lithography step for the maskless fabrication of free-form microfluidic masters is shown in Fig. 1a. The microscopy glass substrates (Sigma-Aldrich, 25 x 75 x 1 mm) used for the microfluidic master fabrication were first cleaned in an ultrasonic bath of isopropanol (Sigma-Aldrich, ≥ 99.8 % (GC)), followed by sonication in acetone (Honeywell, ≥ 99 % (GC)). To promote adequate photoresist adhesion, the glass substrates were silanized overnight by immersion in a solution containing one drop of 3 (trimethoxysilyl)propyl acrylate (Sigma-Aldrich, 92 % with 100 ppm BHT) per 5 mL of acetone.[53] The silanized glass substrates were rinsed with deionized water, blown dry with nitrogen, covered with multiple droplets of photoresist (Nanoscribe, IP-S), and remaining air bubbles were removed with a needle. The photoresist covered substrates were mounted in the multiphoton 3D printer (Nanoscribe, Photonic Professional GT) and the microfluidic masters were printed section-wise with rectangular splitting using a 25x lens with a numerical aperture (NA) of 0.8 to focus the laser beam into the photoresist, inducing two-photon polymerization. To reduce the fabrication time to 6 h, the structure was printed in two sections with an outer shell consisting of 10 layers and an internal tetrahedron scaffold, applying slicing and hatching distances of 400 nm and 200 nm, respectively.[54] After the printing process, the microfluidic masters were developed for 10 min in propylene glycol methyl ether acetate (PGMEA) (Sigma-Aldrich, ≥ 99.5 %) and rinsed for 3 min with isopropanol. The inside of the microfluidic master structure was still composed of uncured photoresist, supported by the polymerized scaffold for structural integrity. The remaining liquid photoresist core was fully solidified in a post-curing step by 12 h exposure to an ultraviolet light source (302 nm, 8 W). A scanning electron microscopy (SEM) (Hitachi, TM3030) image of the resulting microfluidic master is shown in Fig. 2a.

**Mould replication of the microfluidic channels by soft lithography**

Fig. 1b shows the poly(dimethylsiloxane) (PDMS) (Dow Corning, Sylgard® 184 plus curing agent, 10:1 (w/w)) molding and plasma bonding steps.[55] PDMS was cast onto the photoresist master and cured overnight at 60°C in an oven. Holes for the connecting tubing were punched into the detached PDMS slab. Residuals on the PDMS slab were removed by sonication with isopropanol followed by 3 h of drying at atmospheric conditions. To form the microfluidic chip the PDMS slab was plasma bonded to a microscope cover slide (Sigma-Aldrich, 24 x 50 x 0.145 mm). Plasma treatment of the PDMS slab and the glass slide was carried out simultaneously under an absolute oxygen pressure of 120 Pa (40 mL min$^{-1}$) at 40 W for 30 s (TePla 100 Plasma System, PVA).[56] Using a microscope cover slide with a thickness of 145 ± 15 µm ensured that the following in-chip laser lithography step was within the working distance of the microscope. The SEM image in Fig. 2b shows the PDMS nozzle holder of the microfluidic chip before plasma bonding.

**In-chip fluid dynamics simulations**

Computational fluid dynamics (CFD) simulations of the spinneret geometry were carried out with COMSOL Multiphysics® prior to the in-chip fabrication process to gain a better understanding of the fluid dynamic behaviour and to design a spinneret geometry with homogeneous flow around the orifice. The geometry was imported from a computer-aided design (CAD) model using Autodesk Inventor. As a first estimate, a stationary single-phase flow was simulated using the fluid properties of water at standard temperature from the software database. Fully developed laminar inflow conditions and zero outlet pressure were assumed. The mesh was refined towards an average element quality of 0.66 with approximately two million mesh elements leading to a stable pressure-loss probed between the inlet and outlet. Based on the CFD simulation results the spinneret geometry was altered using CAD to influence and optimize towards the desired homogeneous flow field at the nozzle orifice.

**In-chip laser lithography fabrication of the embedded spinneret**

The in-chip laser lithography process used to embed the optimized spinneret structures into the microfluidic PDMS chip is shown in Fig. 1c. As preparation for this printing process, the microfluidic channels were filled with liquid photoresist (Nanoscribe, IP-L). After the alignment and positioning of the coordinate system to the nozzle holder, the spinneret was printed top down with the 25x lens within 30 min, using slicing and hatching distances of 400 nm and 200 nm, respectively. Tubing (Smiths Medical, fine bore polythene, inner diameter: 0.38 mm, outer diameter: 1.09 mm) was attached and glued (UHU, Max Repair Extreme) to the punched holes of the microfluidic chip. The embedded spinneret was then developed by manually flushing acetonitrile (ACN) (Sigma-Aldrich, anhydrous 99.8 %) through the channels for 10 min to remove uncured photoresist. Subsequently, the microfluidic channels were flushed with isopropanol for 3 min and dried with air. Due to the strong swelling of PDMS in PGMEA, ACN was found to be a more appropriate solvent for photoresist development in this system.[57] To improve sealing and structural integrity of the embedded spinneret, silane-based glue (UHU Max Repair Extreme) was pushed into the two sides as shown in Fig. 1c in dark blue. As demonstrated in Fig. 4a, the embedded spinneret was printed into an open microfluidic PDMS channel, which was not bond to a glass substrate by the plasma treatment to obtain SEM images.

**In-chip colour experiments**

The working principle of the embedded spinneret was visualized using two coloured fluids as shown by the optical microscopy image (Motic, AE2000) in Fig. 5a. Blue (E133) and yellow (E100) food colouring (Dr. Oetker GmbH) were separately dispersed in water by ultrasonication at a water to dye ratio of 1:2 (w/w). A system comprising two pressurized vessels was connected to the inlets of the microfluidic chip via the glued tubing. The pressure was adjusted manually until a stable flow profile developed.

**In-chip fibre syntheses**

The embedded spinneret was then used to fabricate micrometre-sized polymer fibres using an inner bore fluid containing 2 % (w/w) polyacrylonitrile (PAN) (Sigma-Aldrich, powder 50 µm mean particle size) dissolved in dimethylsulfoxide (DMSO) (Sigma-Aldrich, ≥ 99.9 %) and an outer sheath fluid containing 35 % (w/w) deionized water (Milli-Q) in DMSO.[48] During the fibre synthesis experiments, the optical microscope including the microfluidic chip was tilted vertically and the chip was dipped into a petri dish filled with the sheath fluid. This ensured that the spun fibre was collected without an air gap at the bottom of the sliced microfluidic chip. The fibre spinning process was initialized by increasing the sheath pressure until a stable flow developed. The bore pressure was then manually adjusted until first the remaining air and later the bore liquid were released through the nozzle tip. After spinning, the fibres were first analysed in a wet state by optical microscopy as shown in Fig. 6a and b. Following the protocol of Nation et al., the fibres were stepwise dehydrated with ethanol (Sigma-Aldrich, ≥ 99.8 % (GC)) and dried using hexamethyldisilazane (HMDS) (Sigma-Aldrich, ≥ 99.0 % (GC)).[58] The fibres were sputtered (Leica, EM ACE600) with a 6 nm gold/palladium layer (4:1 (w/w)) and analysed by field emission scanning electron microscope (FESEM) (Hitachi, S4800) as shown in Fig. 6c and d.

## Conclusions

In this paper, we demonstrate a novel in-chip fabrication strategy that can be used to interface microscopic fluidic structures with macroscopic channel systems. We show the feasibility to make use of additive fabrication methods to prepare intricate three-dimensional microfluidic features and facilitate interfacing with real-world connection systems. Using this in-chip fabrication method, we established a biomimetic wet-spinning process by miniaturization of an artificial spider-inspired spinneret nozzle that is much smaller than the current state-of-the-art to produce wet-spun monofilaments with unseen dimensions of native spider silk. We predict that further process optimization, such as enhanced flow-focusing and the fabrication of smaller nozzles will allow wet-spinning of fibres with even smaller dimensions, potentially within a few hundred nanometres. Whether the presented wet-spinning process will enable the fabrication of artificial spider silk fibres with enhanced morphology and topology remains to be investigated in the future.

This methodology developed here, with dimensions covering length scales over several orders of magnitude, can be interfaced with other additive manufacturing technologies to augment device functionality and versatility.[59-63] We are confident that the presented technology can amplify the knowledge base of fabrication techniques to miniaturize complex and integrated systems.

## Conflicts of interest

There are no conflicts to declare.

## Acknowledgements

Matthias Wessling acknowledges the support through an Alexander-von-Humboldt Professorship. This project has received funding from the European Research Council (ERC) under the European Union's Horizon 2020 research and innovation program (grant agreement no. 694946). This project was partially performed at the Center for Chemical Polymer Technology CPT, which is supported by the EU and the federal state of North Rhine-Westphalia (grant no. EFRE 30 00 883 02).

## References

1. L. Eadie and T. K. Ghosh, *J. R. Soc., Interface*, 2011, **8**, 761-775.
2. J. M. Gosline, M. E. DeMont and M. W. Denny, *Endeavour*, 1986, **10**, 37-43.
3. P. M. Cunniff, S. A. Fossey, M. A. Auerbach, J. W. Song, D. L. Kaplan, W. W. Adams, R. K. Eby, D. Mahoney and D. L. Vezie, *Polym. Adv. Technol.*, 1994, **5**, 401-410.
4. A. E. Albertson, F. Teulé, W. Weber, J. L. Yarger and R. V. Lewis, *J. Mech. Behav. Biomed. Mater.*, 2014, **29**, 225-234.
5. A. Heidebrecht, L. Eisoldt, J. Diehl, A. Schmidt, M. Geffers, G. Lang and T. Scheibel, *Adv. Mater.*, 2015, **27**, 2189-2194.
6. R. Lewis, *Bioscience*, 1996, **46**, 636-638.
7. A. Lazaris, S. Arcidiacono, Y. Huang, J.-F. Zhou, F. Duguay, N. Chretien, E. A. Welsh, J. W. Soares and C. N. Karatzas, *Science*, 2002, **295**, 472-476.
8. X.-X. Xia, Z.-G. Qian, C. S. Ki, Y. H. Park, D. L. Kaplan and S. Y. Lee, *Proc. Natl. Acad. Sci.*, 2010, **107**, 14059-14063.
9. Z. Lin, Q. Deng, X. Y. Liu and D. Yang, *Adv. Mater.*, 2013, **25**, 1216-1220.
10. S. L. Adrianos, F. Teulé, M. B. Hinman, J. A. Jones, W. S. Weber, J. L. Yarger and R. V. Lewis, *Biomacromolecules*, 2013, **14**, 1751-1760.
11. C. G. Copeland, B. E. Bell, C. D. Christensen and R. V. Lewis, *ACS Biomater. Sci. Eng.*, 2015, **1**, 577-584.
12. S. Lin, S. Ryu, O. Tokareva, G. Gronau, M. M. Jacobsen, W. Huang, D. J. Rizzo, D. Li, C. Staii, N. M. Pugno, J. Y. Wong, D. L. Kaplan and M. J. Buehler, *Nat. Commun.*, 2015, **6**, 6892-6892.
13. M. Andersson, Q. Jia, A. Abella, X.-Y. Lee, M. Landreh, P. Purhonen, H. Hebert, M. Tenje, C. V. Robinson and Q. Meng, *Nat. Chem. Biol.*, 2017.
14. C.-H. Choi, H. Yi, S. Hwang, D. A. Weitz and C.-S. Lee, *Lab Chip*, 2011, **11**, 1477-1483.
15. M. Hu, R. Deng, K. M. Schumacher, M. Kurisawa, H. Ye, K. Purnamawati and J. Y. Ying, *Biomaterials*, 2010, **31**, 863-869.
16. J.-H. Jung, C.-H. Choi, S. Chung, Y.-M. Chung and C.-S. Lee, *Lab Chip*, 2009, **9**, 2596-2602.
17. C. M. Hwang, A. Khademhosseini, Y. Park, K. Sun and S.-H. Lee, *Langmuir*, 2008, **24**, 6845-6851.
18. Y. Jun, E. Kang, S. Chae and S.-H. Lee, *Lab Chip*, 2014, **14**, 2145-2160.
19. M. E. Kinahan, E. Filippidi, S. Köster, X. Hu, H. M. Evans, T. Pfohl, D. L. Kaplan and J. Wong, *Biomacromolecules*, 2011, **12**, 1504-1511.
20. S. Rammensee, U. Slotta, T. Scheibel and A. R. Bausch, *Proc. Natl. Acad. Sci.*, 2008, **105**, 6590-6595.
21. H. Bai, R. Sun, J. Ju, X. Yao, Y. Zheng and L. Jiang, *Small*, 2011, **7**, 3429-3433.
22. M. Sun, Y. Xie, J. Zhu, J. Li and J. C. T. Eijkel, *Anal. Chem.*, 2017, **89**, 2227-2231.
23. W. Denk, J. H. Strickler and W. W. Webb, *Science*, 1990, **248**, 73-76.
24. S. Maruo, O. Nakamura and S. Kawata, *Opt. Lett.*, 1997, **22**, 132-134.
25. G. Kumi, C. O. Yanez, K. D. Belfield and J. T. Fourkas, *Lab Chip*, 2010, **10**, 1057-1060.
26. H.-B. Sun, S. Matsuo and H. Misawa, *Appl. Phys. Lett.*, 1999, **74**, 786-788.
27. V. J. Cadarso, N. Chidambaram, L. Jacot-Descombes and H. Schift, *Microsyst. Nanoeng.*, 2017, **3**, 17017.
28. M. Iosin, T. Scheul, C. Nizak, O. Stephan, S. Astilean and P. Baldeck, *Microfluid. Nanofluid.*, 2011, **10**, 685-690.
29. M. H. Olsen, G. M. Hjortø, M. Hansen, Ö. Met, I. M. Svane and N. B. Larsen, *Lab Chip*, 2013, **13**, 4800-4809.
30. J. Wang, Y. He, H. Xia, L.-G. Niu, R. Zhang, Q.-D. Chen, Y.-L. Zhang, Y.-F. Li, S.-J. Zeng, J.-H. Qin and others, *Lab Chip*, 2010, **10**, 1993-1996.
31. L. Amato, Y. Gu, N. Bellini, S. M. Eaton, G. Cerullo and R. Osellame, *Lab Chip*, 2012, **12**, 1135-1142.
32. D. Wu, S.-Z. Wu, J. Xu, L.-G. Niu, K. Midorikawa and K. Sugioka, *Laser Photonics Rev.*, 2014, **8**, 458-467.

33. B. Xu, W.-Q. Du, J.-W. Li, Y.-L. Hu, L. Yang, C.-C. Zhang, G.-Q. Li, Z.-X. Lao, J.-C. Ni, J.-R. Chu and others, *Sci. Rep.*, 2016, **6**, 19989.
34. C. Zhang, Y. Hu, W. Du, P. Wu, S. Rao, Z. Cai, Z. Lao, B. Xu, J. Ni, J. Li and others, *Sci. Rep.*, 2016, **6**, 33281.
35. D. Wu, Q.-D. Chen, L.-G. Niu, J.-N. Wang, J. Wang, R. Wang, H. Xia and H.-B. Sun, *Lab Chip*, 2009, **9**, 2391-2394.
36. T. W. Lim, Y. Son, Y. J. Jeong, D.-Y. Yang, H.-J. Kong, K.-S. Lee and D.-P. Kim, *Lab Chip*, 2011, **11**, 100-103.
37. Y.-J. Liu, P.-Y. Chen, J.-Y. Yang, C. Tsou, Y.-H. Lee, P. L. Baldeck and C.-L. Lin, *Sens. Mater.*, 2014, **26**, 39-44.
38. N. Spannenburg, H. Offerhaus, D. van den Ende, J. Herek, F. Mugele and others, *J. Micro/Nanolithogr., MEMS, MOEMS*, 2015, **14**, 23503-23503.
39. Y.-J. Liu, J.-Y. Yang, Y.-M. Nie, C.-H. Lu, E. D. Huang, C.-S. Shin, P. Baldeck and C.-L. Lin, *Microfluid. Nanofluid.*, 2015, **18**, 427-431.
40. D. Wu, J. Xu, L.-G. Niu, S.-Z. Wu, K. Midorikawa and K. Sugioka, *Light: Sci. Appl.*, 2015, **4**, e228-e228.
41. M. Focsan, A. M. Craciun, S. Astilean and P. L. Baldeck, *Opt. Mater. Express*, 2016, **6**, 1587-1593.
42. R. K. Flitney, *Seals and sealing handbook*, Elsevier, 2011.
43. US Pat., 20160332365A1, 2016.
44. A.-C. Joel, P. Kappel, H. Adamova, W. Baumgartner and I. Scholz, *Arthropod Struct. Dev.*, 2015, **44**, 568-573.
45. H.-B. Sun, K. Takada, M.-S. Kim, K.-S. Lee and S. Kawata, *Appl. Phys. Lett.*, 2003, **83**, 1104-1106.
46. M. Guney and G. Fedder, *J. Micromech. Microeng.*, 2016, **26**, 105011.
47. Y. Zhang, N. E. Benes and R. G. Lammertink, *Lab Chip*, 2015, **15**, 575-580.
48. J. Chen, C.-g. Wang, X.-g. Dong and H.-z. Liu, *J. Polym. Res.*, 2006, **13**, 515-519.
49. X. Qin, Y. Lu, H. Xiao and W. Zhao, *Polym. Eng. Sci.*, 2013, **53**, 827-832.
50. T. Wang and S. Kumar, *J. Appl. Polym. Sci.*, 2006, **102**, 1023-1029.
51. D. Edie, *Carbon*, 1998, **36**, 345-362.
52. O. Paris, D. Loidl and H. Peterlik, *Carbon*, 2002, **40**, 551-555.
53. K. L. Mittal, *Silanes and Other Coupling Agents*, CRC Press, Boca Raton, 2007.
54. US Pat., 20160114530A1, 2015.
55. Y. Xia and G. M. Whitesides, *Annual Review of Materials Science*, 1998, **28**, 153-184.
56. S. Bhattacharya, A. Datta, J. M. Berg and S. Gangopadhyay, *J. Micromech. Microeng.*, 2005, **14**, 590-597.
57. J. N. Lee, C. Park and G. M. Whitesides, *Anal. Chem.*, 2003, **75**, 6544-6554.
58. J. L. Nation, *Stain Technology*, 1983, **58**, 347-351.
59. S. Z. Guo, F. Gosselin, N. Guerin, A. M. Lanouette, M. C. Heuzey and D. Therriault, *Small*, 2013, **9**, 4090-4090.
60. P. Paiè, F. Bragheri, D. Di Carlo and R. Osellame, *Microsyst. Nanoeng.*, 2017, **3**, 17027.
61. T. Femmer, A. Jans, R. Eswein, N. Anwar, M. Moeller, M. Wessling and A. J. C. Kuehne, *ACS Appl. Mater. Interfaces*, 2015, **7**, 12635-12638.
62. C. C. Glick, M. T. Srimongkol, A. J. Schwartz, W. S. Zhuang, J. C. Lin, R. H. Warren, D. R. Tekell, P. A. Satamalee and L. Lin, *Microsyst. Nanoeng.*, 2016, **2**, 16063.
63. T. G. Leong, A. M. Zarafshar and D. H. Gracias, *Small*, 2010, **6**, 792-806.